\documentclass[sigconf]{acmart}

\usepackage{graphicx}
\usepackage{balance}  % for  \balance command ON LAST PAGE  (only there!)
\usepackage{hyperref}

\usepackage{listings}
\usepackage{subfigure}
\usepackage[linesnumbered,ruled,vlined]{algorithm2e}
\usepackage{algpseudocode}
\usepackage{amsmath}
\usepackage{graphics}
\usepackage{epsfig}
\usepackage{booktabs}
\usepackage{color}
\usepackage{url}
\usepackage{xspace}
\usepackage{verbatim}
\usepackage[export]{adjustbox}

\setcopyright{none}
\acmDOI{10.1145/3205289.3205290}

\acmISBN{978-1-4503-5783-8/18/06}

\acmConference[ICS'18]{The 32nd ACM International Conference on Supercomputing}{June 12--15, 2018}{Beijing, China} 
\acmYear{2018}
\copyrightyear{2018}

\acmPrice{15.00}

\begin{document}

\title{Accurate, Fast and Scalable Kernel Ridge Regression on Parallel and Distributed Systems}
%\titlenote{Produces the permission block, and copyright information}
%\subtitle{Extended Abstract}
%\subtitlenote{The full version of the author's guide is available as \texttt{acmart.pdf} document}

\author{Yang You, James Demmel}
%\authornote{Dr.~Trovato insisted his name be first.}
\orcid{}
\affiliation{%
  \institution{UC Berkeley}
  \streetaddress{}
  \city{} 
  \state{} 
  \postcode{}
}
\email{{youyang, demmel}@cs.berkeley.edu}

\author{Cho-Jui Hsieh}
%\authornote{Dr.~Trovato insisted his name be first.}
\orcid{}
\affiliation{%
  \institution{UC Davis}
  \streetaddress{}
  \city{} 
  \state{} 
  \postcode{}
}
\email{chohsieh@ucdavis.edu}

\author{Richard Vuduc}
%\authornote{Dr.~Trovato insisted his name be first.}
\orcid{}
\affiliation{%
  \institution{Georgia Tech}
  \streetaddress{}
  \city{} 
  \state{} 
  \postcode{}
}
\email{richie@cc.gatech.edu}

\iffalse
\author{Yang You, Cho-Jui Hsieh, Richard Vuduc, James Demmel}
%\authornote{Dr.~Trovato insisted his name be first.}
\orcid{}
\affiliation{%
  \institution{UC Berkeley, UC Davis, Georgia Tech}
  \streetaddress{}
  \city{} 
  \state{} 
  \postcode{}
}
\email{{youyang, demmel}@cs.berkeley.edu, chohsieh@ucdavis.edu, richie@cc.gatech.edu}
\fi
%\author{G.K.M. Tobin}
%\authornote{The secretary disavows any knowledge of this author's actions.}
%\affiliation{%
  %\institution{Institute for Clarity in Documentation}
  %\streetaddress{P.O. Box 1212}
  %\city{Dublin} 
  %\state{Ohio} 
  %\postcode{43017-6221}
%}
%\email{webmaster@marysville-ohio.com}

%\author{Lars Th{\o}rv{\"a}ld}
%\authornote{This author is the
  %one who did all the really hard work.}
%\affiliation{%
  %\institution{The Th{\o}rv{\"a}ld Group}
  %\streetaddress{1 Th{\o}rv{\"a}ld Circle}
  %\city{Hekla} 
  %\country{Iceland}}
%\email{larst@affiliation.org}

% The default list of authors is too long for headers}
%\renewcommand{\shortauthors}{B. Trovato et al.}

\begin{abstract}
Kernel Ridge Regression (KRR) is a fundamental method in machine learning.
Given an $n$-by-$d$ data matrix as input, a traditional implementation requires $\Theta(n^2)$ memory to form an $n$-by-$n$ kernel matrix and $\Theta(n^3)$ flops to compute the final model.
These time and storage costs prohibit KRR from scaling up to large datasets.  
For example, even on a relatively small dataset
(a 520k-by-90 input requiring 357~MB), 
KRR requires 2~TB memory just to store the kernel matrix. 
The reason is that $n$ usually is much larger than $d$ for real-world applications.
On the other hand, weak scaling becomes a problem:
if we keep $d$ and $n/p$ fixed as $p$ grows ($p$ is \# machines), 
the memory needed grows as $\Theta(p)$ per processor and the flops as $\Theta(p^2)$ per processor.
In the perfect weak scaling situation, both the memory needed and the flops grow as $\Theta(1)$ per processor (i.e. memory and flops are constant).
The traditional Distributed KRR implementation (DKRR) only achieved 0.32\% weak scaling efficiency from $96$ to $1536$ processors.

We propose two new methods to address these problems: the Balanced KRR (BKRR) and K-means KRR (KKRR).
These methods consider alternative ways to partition the input dataset into $p$ {\bf different} parts, generating $p$ {\bf different} models, 
and then selecting the {\bf best} model among them.
Compared to a conventional implementation, KKRR2 (optimized version of KKRR) improves the weak scaling efficiency from 
0.32\% to 38\% and achieves a 591$\times$ speedup for getting the same accuracy by using the same data and the same hardware (1536 processors).
BKRR2 (optimized version of BKRR) achieves a higher accuracy than the current fastest method using less training time for a variety of datasets.
For the applications requiring only approximate solutions, BKRR2 improves the weak scaling efficiency to 92\% and achieves 3505$\times$ speedup (theoretical speedup: 4096$\times$).
%The source code is available online\footnote{\label{footnote:code}The source code (anonymous link for review) of this paper is at \href{https://www.dropbox.com/s/p63vvuylezi52y0/krr.zip}{https://www.dropbox.com/s/p63vvuylezi52y0/krr.zip}}.
\end{abstract}

%
% The code below should be generated by the tool at
% http://dl.acm.org/ccs.cfm
% Please copy and paste the code instead of the example below. 
%

% We no longer use \terms command
%\terms{Theory}


\maketitle

\section{Introduction}

Learning non-linear relationships between predictor variables and responses is a fundamental problem in machine learning~\cite{barndorff2004econometric}, \cite{bertin2011million}.
One state-of-the-art method is Kernel Ridge Regression (KRR)~\cite{zhang2013divide}, which we target in this paper.
It combines ridge regression, a method to address ill-posedness and overfitting in the standard regression setting via L2 regularization, with kernel techniques, which adds flexible support for capturing non-linearity.
%[RV] This stuff is fluffy:
%A fundamental problem of statistical machine learning is 
%Regression is an important machine learning technique that has been widely used in many fields such as
%recommendation systems \cite{bertin2011million} and
%Financial Economics \cite{barndorff2004econometric}. To solve ill-posed problems and
%avoid overfitting, Ridge Regression was designed to add L2 regularization to linear regression.
%Furthermore, to learn a non-linear relationship between input data and responses, it is common to combine the Kernel method with Ridge Regression. 
%The resulting Kernel Ridge Regression (KRR) is a state-of-the-art regression algorithm. 

Computationally, the input of KRR is an $n$-by-$d$ data matrix with $n$ training samples and $d$ 
features, where typically $n\gg d$. % for big data applications.
At training time,
KRR needs to solve a large linear system $(K+\lambda n I) \alpha = y$
where $K$ is an $n$-by-$n$ matrix, $\alpha$ and $y$ are $n$-by-1 vectors, and $\lambda$ is a scalar.
Forming and solving this linear system is the major bottleneck of KRR. For 
example, even on a relatively small
dataset---357~megabytes (MB) for a 520,000-by-90~\cite{bertin2011million}---KRR needs to form a 2~terabyte dense kernel matrix.
A standard distributed parallel dense linear solver on a $p$-processor system will require $\Theta(n^3/p)$ arithmetic operations per processor;
we refer to this approach hereafter as Distributed KRR (DKRR).
In machine learning, where weak scaling is of primary interest, DKRR fares poorly:
keeping $n/p$ fixed, the total storage grows as $\Theta(p)$ per processor
and the total flops as $\Theta(p^2)$ per processor.
In the perfect weak scaling situation, both the memory needed and the flops grow as $\Theta(1)$ per processor (i.e. memory and flops are constant).
In one experiment, the weak scalability of DKRR dropping from 100\% to 0.32\% as $p$ increased from $96$ to $1536$ processors.

Divide-and-Conquer KRR (DC-KRR)~\cite{zhang2013divide}, addresses the scaling problems of DKRR.
Its main idea is to randomly shuffle the rows of the $n$-by-$d$ data matrix and then partition it
to $p$ different $n/p$-by-$d$ matrices, one per machine, leading to an
$n/p$-by-$n/p$ kernel matrix on each machine;
it builds local KRR models that are then reduced globally to obtain the final model.
%DC-KRR outperforms DKRR by reducing the per-processor memory requirement from $\Theta(n^2/p)$ to $\Theta(n^2/p^2)$ and per-processor flop count from $\Theta(n^3/p)$ to $\Theta(n^3/p^3)$.
DC-KRR reduced the memory and computational requirement.
%It is generally much faster than DKRR.
However, it can't be used in practice because it sacrifices accuracy.
%However, the accuracy of DC-KRR is often not satisfactory due to the random 
%partition of training data. 
%{\color{red} For example, it can only achieve xxx 
  %MSE (Mean Square Error) on the MSD dataset with xxx samples, while the 
  %original KRR can achieve xxx MSE. 
 %} 
%Firstly, our method is much more accurate than DC-KRR (Fig. \ref{fig:dckrr_cakrr_example}). 
For example, on the Million Song Dataset (a recommendation system application) with 2k test samples, the mean squared error (MSE) decreased only from 88.9 to 81.0 
as the number of training samples increased from 8k to 128k.
We double the number of processors as we double the number of samples. This is a bad weak scaling problem in terms of accuracy.
By contrast, the MSE of DKRR decreased from 91 to 0.002, which is substantially better.

\begin{figure}[!t]
\centering
\includegraphics[width=3.5in]{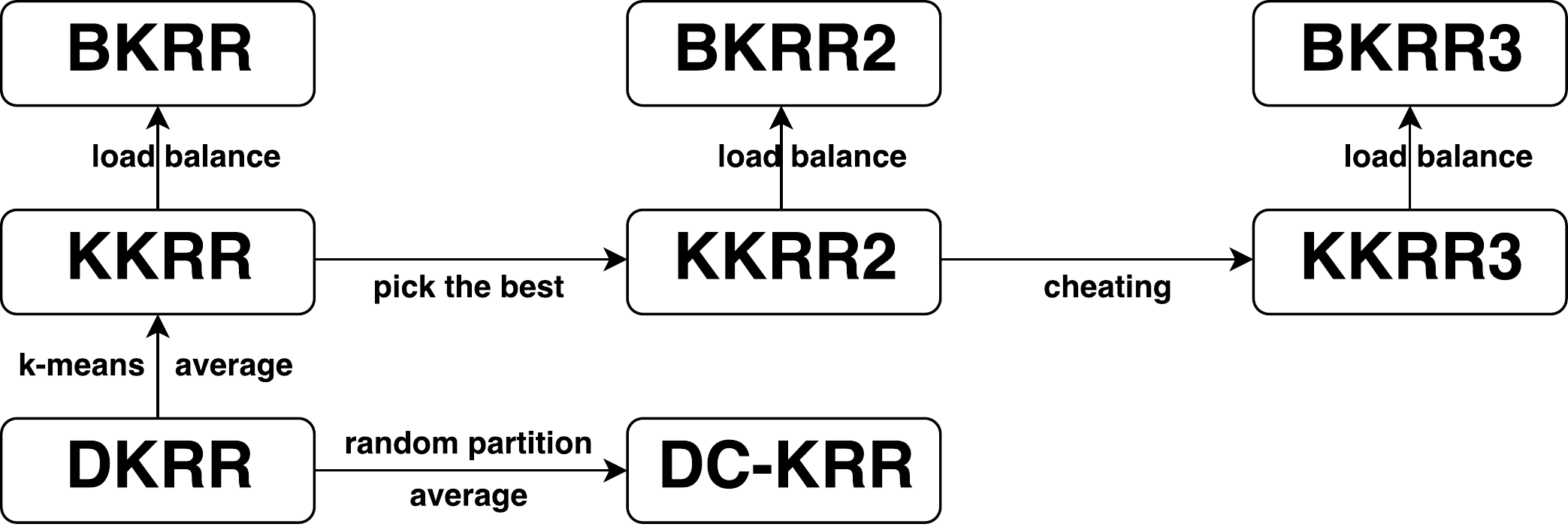}
\caption{Optimization flow of our algorithm. DKRR is the baseline. DC-KRR is the existing method. All the others are the new approaches proposed in this paper.}
\label{fig:optimization_flow}
\end{figure}

\begin{figure}[!t]
\centering
\includegraphics[width=3.3in]{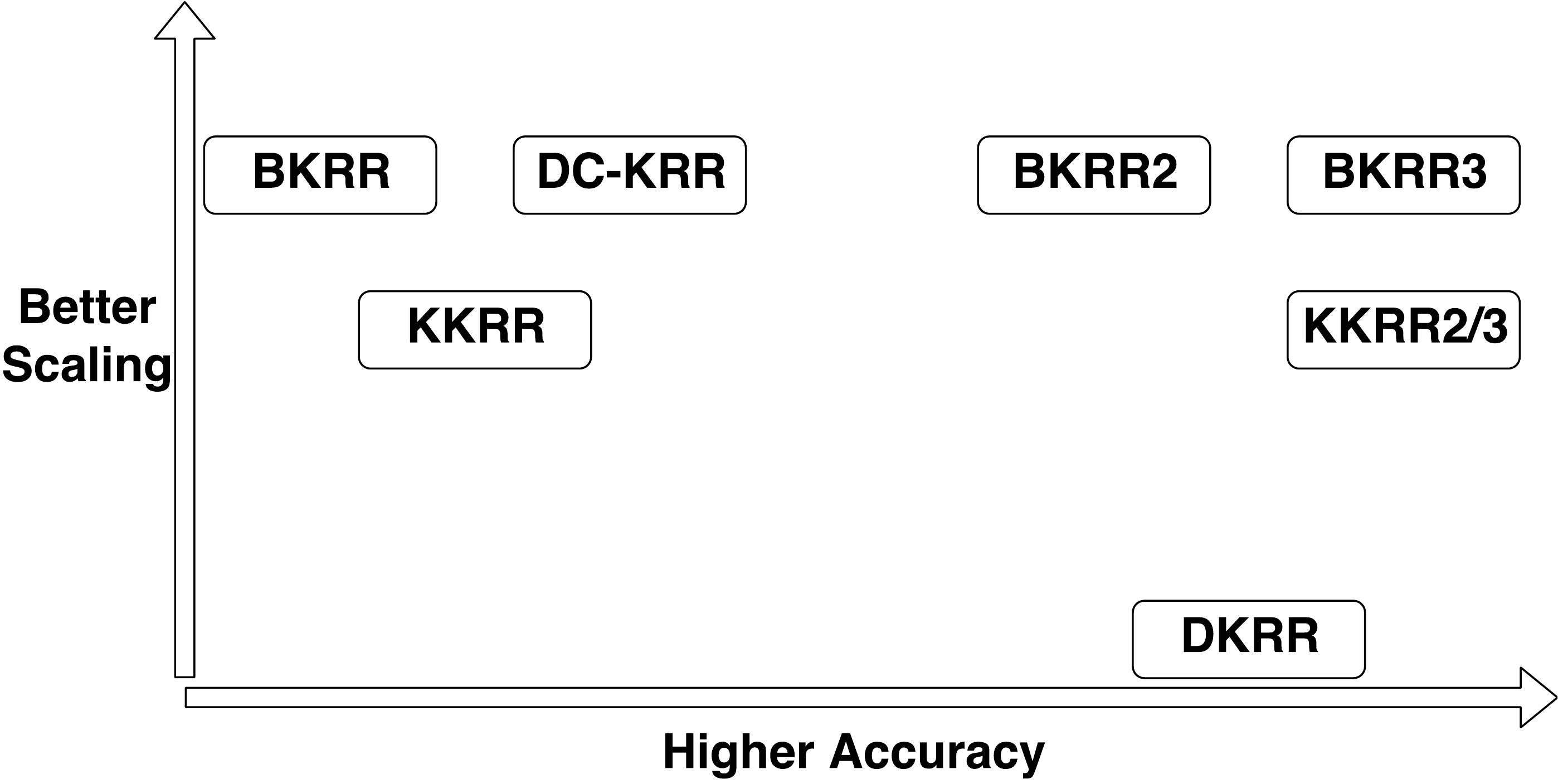}
\caption{Trade-off between accuracy and speed for large-scale weak scaling study. BKRR3 and KKRR3 are the unrealistic approaches. BKRR2 is optimal for scaling and has good accuracy. KKRR2 is optimal for accuracy and has good speed.}
\label{fig:krr_comparison}
\end{figure}

The idea of DC-KRR is to partition the dataset into $p$ {\bf similar} parts
and generate $p$ {\bf similar} models, and then  
{\bf average} these $p$ models to get the final solution.
We seek other ways to partition the problem that scale as well as DC-KRR while improving its accuracy.
Our idea is to partition the input dataset into $p$ {\bf different} parts and generate $p$ {\bf different} models, from which we then select the {\bf best} model among them.
Further addressing particular communication overheads, we obtain two new methods, which we call Balanced KRR (BKRR) and K-means KRR (KKRR).
%We show this strategy gives much higher accuracy than DC-KRR using the same training time.
Figure \ref{fig:optimization_flow} is the summary of our optimization flow. We proposed a series of approaches with details explained later (KKRR3 is an impractical algorithm used later to bound the attainable accuracy).
Figure \ref{fig:krr_comparison} shows the fundamental trade-off between accuracy and scaling for these approaches. 
Among them, we recommend BKRR2 (optimized version of BKRR) and KKRR2 (optimized version of KKRR) to use in practice. BKRR2 is optimized for scaling and has good accuracy. KKRR2 is optimized for accuracy and has good speed.
%Figure \ref{fig:cakrr_dckrr_diff} shows the difference between BKRR2 and DC-KRR.

When we increase the number of samples from 8k to 128k, KKRR2 (optimized version of KKRR) reduces the MSE from 95 to $10^{-7}$, which addresses the poor accuracy of DC-KRR.
In addition, KKRR2 is faster than DC-KRR for a variety of datasets. %DC-KRR can never get the best accuracy achieved by CA-KRR.
Our KKRR2 method improves weak scaling efficiency over DKRR from 0.32\% to 38\% and achieves a 591$\times$ speedup over DKRR on the same data at the same accuracy and the hardware (1536 processors).
For the applications requiring only approximate solutions, BKRR2 improves the weak scaling efficiency to 92\% and achieves 3505$\times$ speedup with only a slight loss in accuracy.
%The source code for our methods is available$^{\ref{footnote:code}}$. %at \href{http://www.cs.berkeley.edu/~youyang/cakrr.zip}{here}.

\section{Background}
%We introduce Kernel Ridge Regression (KRR) in this section, 
%which is a combination of ridge regression with the kernel method. Also, we 
%introduce some background of standard K-means clustering, which will be used 
%as a baseline for partitioning data points.  
\subsection{Linear Regression and Ridge Regression}
In machine learning, linear regression is a widely-used method for modeling the relationship
between a scalar dependent variable $y$ (regressand) and multiple explanatory variables (independent variables)
denoted by $x$, where $x$ is a $d$-dimensional vector.
The training data of a linear regression problem consists of $n$ data points, where each data point (or training sample) is a pair $(x_i, y_i)$
and $1 \leq i \leq n$.
Linear regression has two phases: training and prediction. The training phase builds a model
from an input set of training samples, while the prediction phase uses the model to predict the
unknown regressands $\hat{y}$ of a new data point $\hat{x}$.  The training phase is the main limiter to scaling,
both with respect to increasing the training set size $n$ and number of processors $p$.
In contrast, prediction is embarrassingly parallel and cheap per data point.
%Therefore, this paper focuses on training

{\bf Training Process.}
Given $n$ training data points $\{(x_i, y_i)\}_{i=1}^n$ where each 
$x_i = (x_i^1, ..., x_i^j, ..., x_i^d)$ and $y_i$ is a scalar, 
we want to build a model relating
a given sample ($x_i$) to its measured regressand ($y_i$). 
For convenience, we define $X$ as an $n$-by-$d$ matrix with $X_{ij}=x^j_i$ 
and $y=(y_1, \cdots, y_n)$ be an $n$-dimensional vector. 
The goal of regression is to find a $w$ such that $y_i\approx w^T \cdot x_i$.
%Let us define a $d$-dimensional vector of unknown $w$,
%$X$ is a $n$-by-$d$ matrix with $X_{ij} = x_i^j$, and $y$ is an $n$-dimensional vector.
% The $i$-th row of $X$ is the $i$-th sample and $i$-th element of $y$ is the $i$-th regressand ($i \in \{1, 2, ..., n\}$).
%We want $b$ to satisfy $y_i \approx b^T \cdot x_i$.
This can be formulated as a least squares problem in Equation~(\ref{equation:liner_regression}).
To solve some ill-posed problems or prevent overfitting, L2 regularization \cite{schmidt2005least}
is used, as formulated in Equation~(\ref{equation:ridge_regression}), which is called {\bf Ridge Regression}.
$\lambda$ is a positive parameter that controls $w$: the larger is $\lambda$, the smaller is $\|w\|_2$. 

\begin{equation}
\small
{argmin}_w \; \| Xw - y \|_2^2
\label{equation:liner_regression}
\end{equation}

\begin{equation}
\small
{argmin}_w \; \{ \| Xw - y \|_2^2 + \lambda \|w\|_2^2\}
\label{equation:ridge_regression}
\end{equation}

{\bf Prediction Process.}
Given a new sample $\hat{x}$, 
%and its true regressand ($\hat{y}$),
we can use the $w$ computed in the training process to predict the regressand 
of $\hat{x}$ by
%$\tilde{y}$ by 
%of $\hat{x}$:
$\tilde{y}= w^T \cdot \hat{x}$. If we have $k$ test samples $\{\hat{x}_i\}_{i=1}^k$ 
and their true regressands $\{\hat{y}_i\}_{i=1}^k$, the accuracy of the 
estimated regressand $\{\tilde{y}_i\}_{i=1}^n$ 
can be evaluated by MSE (Mean Squared Error) defined in Equation (\ref{equation:MSE}). %Lower MSE means the method has a higher accuracy.

\begin{equation}
\small
MSE = \frac{1}{k} {\sum}_{i=1}^k {(\tilde{y}_i - \hat{y}_i)}^2
\label{equation:MSE}
\end{equation}

\subsection{Kernel Method}
For many real problems, the underlying model cannot be described by a linear function, so linear ridge regression suffers from poor prediction error. In those cases, a common approach is to map samples to a high dimensional space using a nonlinear mapping, and then learn the model in the high dimensional space. Kernel method \cite{hofmann2008kernel} is a widely used approach to conduct this learning procedure implicitly by defining the kernel function---the similarity of samples in the high dimensional space. 
%In many cases the model may not be a linear function. In thoses cases, 
%kernel methods have to be used to map low dimensional data to higher dimensions in order to get
%a better understanding of nonlinear properties of the original data \cite{hofmann2008kernel}.
%After kernel mapping, the dimension of the dataset may be too high to represent.
%% For example, after Gaussian kernel mapping, the dimensional can be infinite.
%The kernel function allows the user to use the kernel method by defining the similarity between samples in the high dimensional space. 
The commonly used kernel functions are shown in Table~\ref{table:kernels}, 
%of which the Gaussian kernel is the most widely used. We use Gaussian kernel in this paper.
and we use the most widely used Gaussian kernel in this paper. 

\begin{table}[!t]
\footnotesize
\renewcommand{\arraystretch}{1.3}
\caption{Standard Kernel Functions}
\centering
\begin{tabular}{l*{6}{c}r}
\hline
Linear & $\Phi(x_i,x_j) = {x_i}^\top x_j$ \\
\hline
Polynomial & $\Phi(x_i,x_j) = (a {x_i}^\top x_j + r)^d$ \\
\hline
Gaussian & $\Phi(x_i,x_j) = \exp(-||x_i - x_j||^2/(2  {\sigma}^2))$ \\
\hline
Sigmoid & $\Phi(x_i,x_j) = \tanh(a {x_i}^\top x_j + r)$ \\
%\hline
%Laplacian & $\Phi(x_i, x_j) = \exp({-\gamma ||X_i-X_j||_1})$ \\
\hline
\end{tabular}
\label{table:kernels}
\end{table}

\subsection{Kernel Ridge Regression (KRR)}
Combining the Kernel method with Ridge Regression yields Kernel Ridge Regression,
which is presented in Equations~(\ref{equation:krr}) and (\ref{equation:fi}).
The $\|\cdot\|_H$ in Equation (\ref{equation:krr}) is a  Hilbert space norm \cite{zhang2013divide}.
Given the $n$-by-$n$ kernel matrix $K$, this problem reduces to a linear system defined in Equation~(\ref{equation:alpha}).
$K$ is the kernel matrix constructed from training data by $K_{i,j}=\Phi(x_i,x_j)$,
$y$ is the input $n$-by-1 regressand vector corresponding to $X$, and $\alpha$ is the $n$-by-1 unknown solution vector.

\begin{equation}
\small
argmin \; {1 \over n}{\sum}_{i=1}^{n}\| f_i - y_i \|_2^2 + \lambda \|f\|_H^2
\label{equation:krr}
\end{equation}

\begin{equation}
\small
f_i = {\sum}_{j=1}^{n} {\alpha}_j \Phi(x_j, x_i)
\label{equation:fi}
\end{equation}

%\textcolor{blue}{********************************************** I stopped here **********************************************}
In the Training phase, the algorithm's goal is to get $\alpha$ by (approximately) solving the linear system in (\ref{equation:alpha}).
In the Prediction phase, the algorithm uses $\alpha$ to predict the regressand of any unknown sample $\hat{x}$ using Equation (\ref{equation:predict}). 
KRR is specified in Algorithm~\ref{algo:krr}.
The algorithm searches for the best $\sigma$ and $\lambda$ from parameter sets. 
Thus, in practice, Algorithm~\ref{algo:krr} is only a single iteration of KRR because people do not know the best parameters before using the dataset. 
$|\Lambda| \times |\Sigma|$ is the number of iterations where $\Lambda$ and $\Sigma$ are the parameter sets of $\lambda$ and $\sigma$ (Gaussian Kernel) respectively. Thus, the computational cost of KRR method is $\Theta(|\Lambda||\Sigma|n^3)$. In a typical case, if $|\Lambda| = 50$ and $|\Sigma| = 50$, then the algorithm needs to finish thousands of iterations. 
People also use cross-validation technique to select the best parameters, which needs much more time. 
%If we call lines 3-12 as one epoch, then the cross-validation needs to finish $e$ epochs. Typically, $e$ can be 100 or 1000. The dataset will be reordered in each epoch.

\begin{equation}
\small
(K + \lambda n I) \alpha = y
\label{equation:alpha}
\end{equation}

\begin{equation}
\small
\tilde{y} = {\sum}_{i=1}^{n} {\alpha}_i \Phi(x_i, \hat{x})
\label{equation:predict}
\end{equation}

\begin{algorithm}
\small
\DontPrintSemicolon 
\KwIn{\newline
$n$ labeled data points $(x_i, y_i)$ for training;\newline
another $k$ labeled data points $(\hat{x}_j, \hat{y}_j)$ for testing; \newline
both $x_i$ and $\hat{x}_j$ are $d$-dimensional vectors; \newline
$i \in \{1, 2, ..., n\}$, $j \in \{1, 2, ..., k\}$; \newline
tuned parameters $\lambda$ and $\sigma$}
\KwOut{\newline Mean Squared Error ($MSE$) of prediction}
Create a $n$-by-$n$ kernel matrix $K$ \;
	\For{$i \in 1:n$} {
    		\For{$j \in 1:n$} {
        			$K[i][j]$ $\leftarrow$ $\Phi$($x_i$, $x_j$) based on Table \ref{table:kernels}\;
    		}
	}
Solve linear equation $(K+\lambda n I) \alpha = y$ for $\alpha$\;
\For{$j \in 1:k$} {
    $\tilde{y}_j \leftarrow {\sum}_{i=1}^{n} {\alpha}_i K(x_i, \hat{x}_j)$\;
}
$MSE \leftarrow \frac{1}{k} {\sum}_{j=1}^{k} (\tilde{y}_j  - \hat{y}_j)^2$\;
\caption{Kernel Ridge Regression (KRR)}
\label{algo:krr}
\end{algorithm}

\subsection{K-means clustering}
Here we review K-means clustering algorithm, which will be used in our algorithm.  
The objective of K-means clustering is to partition a dataset $TD$ into $k \in Z^+$ subsets
($TD_1, TD_2, ..., TD_k$), using a notion of proximity based on Euclidean distance~\cite{forgy1965cluster}.
The value of $k$ is chosen by the user. Each subset has a center ($CT_1, CT_2, ..., CT_k$), each of which is a $d$-dimensional vector.
%\COMMENT{Is this really true, that the center is a sample and not the center-of-mass?}
A sample $x$ belongs to $TD_i$ if $CT_i$ is its closest center.
K-means is shown in Algorithm~\ref{algorithm:kmeans}.

\begin{algorithm}[htb]
\small
\caption{Plain K-means Clustering}
\label{algorithm:kmeans}
%{\bf 0}:
%Start
%{\bf 1}:
Input the training samples $x_i$, $i \in \{1,2,...,n\}$

%{\bf 2}:
Initialize data centers $CT_1, CT_2, ..., CT_k$ randomly

%{\bf 3}:
$\delta$ $\leftarrow$ $0$

%{\bf 4}:
For every $i=1, \cdots, n$
 
--- $c^{i}$ $\leftarrow$ ${argmin}_j||x_i - CT_j||$

%{\bf 5}:
--- If $c^{i}$ has been changed, $\delta$ $\leftarrow$ $\delta + 1$

End For

%{\bf 6}:
For every $j=1, \cdots, k$ 

--- $CT_j$ $\leftarrow$ $\frac{\sum_{i=1}^{n}1\{c^i=j\}x_i}{\sum_{i=1}^{n}1\{c^i=j\}}$%, $j \in \{1,2,...,k\}$

End For

%{\bf 7}:
If $\delta/n$ $>$ threshold, then go to Step 3

%{\bf 12}:
%End
\label{algorithm:K-means}
\end{algorithm}

\section{Existing Methods}
\subsection{Distributed KRR (DKRR)}
\begin{figure}[!t]
\centering
\includegraphics[width=2.7in]{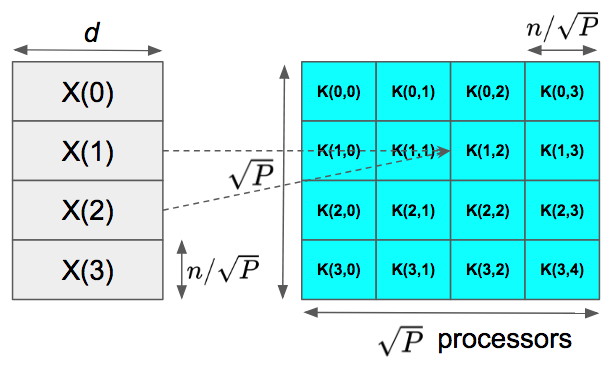}
\caption{Implementation of Distributed KRR. We divide the input sample matrix into $\sqrt{p}$ parts by rows, and each machine gets two of these $\sqrt{p}$ parts. Then each machine generates $1/p$ part of the kernel matrix.}
\label{fig:diskrr}
\end{figure}

The bottleneck of KRR is solving the $n$-by-$n$ linear system (\ref{equation:alpha}),
which is generated by a much smaller $n$-by-$d$ input matrix with $n \gg d$.
As stated before, this makes weak-scaling problematic, because
memory-per-machine grows like $\Theta (p)$, and
flops-per-processor grows like $\Theta (p^2)$.
In the perfect weak scaling situation, both the memory needed and the flops grow as $\Theta(1)$ per processor (i.e. memory and flops are constant).
%For example, a 357 MB dataset (520,000 $\times$ 90 matrix) \cite{bertin2011million} would generate a 2000 GB kernel matrix.
Our experiments show the weak scaling efficiency of DKRR decreases from 100\% to 0.3\% as we increase the number of processors from 96 to 1536.
Since the $n$-by-$n$ matrix $K$ cannot be created on a single node, we create it in a distributed way
on a $\sqrt{p}$-by-$\sqrt{p}$ machine grid (Fig. \ref{fig:diskrr}).
We first divide the sample matrix into $\sqrt{p}$ equal parts by rows.
To generate $\frac{1}{p}$ of the kernel matrix, each machine will need two of these $\sqrt{p}$ parts of the sample matrix.
For example, in Fig. \ref{fig:diskrr}, to generate the $K(1,2)$ block,
machine~6 needs the second and the third parts of the blocked sample matrix.
Thus, we reduce the storage and computation for kernel creation
from $\Theta(n^2)$ to $\Theta(n^2/p)$ per machine.
Then we use distributed linear solver in ScaLAPACK \cite{choi1995scalapack}
to solve for $\alpha$.

%\textcolor{blue}{Line 3 of Algorithm~\ref{algo:krr} is for cross-validation, please also update the line pointer of this part}

\subsection{Divide-and-Conquer KRR (DC-KRR)}
DC-KRR \cite{zhang2013divide} showed that using divide-and-conquer can reduce the memory and computational requirement. 
%We implement this serial approach on distributed systems. 
We first shuffle the original data matrix
$M = [X, y]$ by rows. Then we distribute $M$ to all the nodes evenly by rows (lines 1-5 of Algorithm~\ref{algo:rkrr}).
On each node, we construct a much smaller matrix ($\Theta(n^2/p^2)$) than the original kernel matrix ($\Theta(n^2)$)
(lines 6-11 of Algorithm~\ref{algo:rkrr}). After getting the local kernel matrix $K$, we use it to solve a
linear equation $(K+\lambda n I) \alpha = y$ on each node where $y$ is the input labels and
$\alpha$ is the solution. After getting $\alpha$, the training step is complete.
Then we use the local $\alpha$ to make predictions for each unknown data point and
do a global average (reduction) to output the final predicted label (lines 13-15 of Algorithm~\ref{algo:rkrr}).
%We got a Matlab code from the authors of \cite{zhang2013divide} and implemented it on distributed systems. 
%Our implementation got the same result as their implementation for the same input and the same parameters.
If we get a set of better hyper-parameters, we record them by overwriting the old versions (lines 16-19 of Algorithm~\ref{algo:rkrr}).

\begin{algorithm}
\small
\DontPrintSemicolon % Some LaTeX compilers require you to use \dontprintsemicolon instead
\KwIn{\newline$n$ labeled data points $(x_i, y_i)$ for training;\newline
$k$ labeled data points $(\hat{x}_j, \hat{y}_j)$ for testing;\newline
$x_i$ and $\hat{x}_j$ are $d$-dimensional vectors;\newline
$i \in \{1, 2, ..., n\}$, $j \in \{1, 2, ..., k\}$;\newline
$\widehat{MSE}$ $\leftarrow$ $\infty$ (Initial Mean Squared Error);}
\KwOut{\newline Mean Squared Error ($\widehat{MSE}$) of prediction; \newline best parameters $\hat{\lambda}, \hat{\sigma}$;}
\If{$rank = 0$} {
	Store data points $(x_i, y_i)$ $i \in \{1, ..., n\}$ as [$X$, $y$]\;
	$M$ = [$X$, $y$] is a $n$-by-$(d+1)$ matrix\;
	Shuffle $M$ by rows\;
	Scatter $M$ to all the nodes evenly by rows\;
}
Create a $m$-by-$m$ kernel matrix $K$, $m = n/p$\newline
\For{$i \in 1:m$} {
	$x_i, y_i$ = M[i][1:d+1]
}
\For{$\lambda \in \Lambda$ and $\sigma \in \Sigma$} {
\For{$i \in 1:m$} {
    \For{$j \in 1:m$} {
        $K[i][j]$ = $\Phi$($x_i$, $x_j$) based on Table \ref{table:kernels}\;
    }
}
Solve linear equation $(K+\lambda m I) \alpha = y$ for $\alpha$\;
\For{$j \in 1:k$} {
    $\bar{y}_j = {\sum}_{i=1}^{m} {\alpha}_i K(x_i, \hat{x}_j)$\;
}
Global reduce: $\tilde{y} = \sum \bar{y}/p$\;
\If{$rank = 0$} {
	$MSE = \frac{1}{k} {\sum}_{j=1}^{k} (\tilde{y}_j  - \hat{y}_j)^2$\;
	\If{$MSE < \widehat{MSE}$} {
    		$\widehat{MSE} \leftarrow MSE$, $\widehat{\lambda} \leftarrow \lambda$, $\widehat{\sigma} \leftarrow \sigma$\;
	}
}
}
\caption{Divide-and-Conquer KRR (DC-KRR)}
\label{algo:rkrr}
\end{algorithm}

\section{Accurate, Fast, and Scalable KRR}
\subsection{Kmeans KRR (KKRR)}
DC-KRR performs better than state-of-the-art methods \cite{zhang2013divide}.
However, based on our observation, DC-KRR still needs to be improved. DC-KRR has a poor weak scaling in terms of accuracy, which is the bottleneck for distributed machine learning workloads.
%We want to get the highest accuracy by the fastest approach.
%In fact, the approximation error can be further reduced (more details in Section \ref{sec:appro}).
% We want to design the method that is not only fast enough but also can approximate the exact solution as close as possible.
We observe that the poor accuracy of DC-KRR is mainly due to the random 
partitioning of training samples. Thus, our objective is to design a better partitioning method to achieve a higher accuracy and maintain a high scaling efficiency. 
Our algorithm is accurate, fast, and scalable on distributed systems. 
%Here, we show that using K-means to partition the original data.

The analysis in this section is based on the Gaussian kernel because it is the
most widely used case \cite{zhang2013divide}. Other kernels can work in the same way with different distance metrics.
For any two training samples, their kernel value ($exp\{-\|x_i - x_j\|^2/(2{\sigma}^2)\}$) is close to zero when their Euclidean distance ($\|x_i - x_j\|^2$) is large. Therefore, for a given sample $\hat{x}$, only the training points close to
$\hat{x}$ in Euclidean distance can have an effect on the result (Equation~(\ref{equation:predict})) of the prediction process.
Based on this idea, we partition the training dataset into $p$ subsets ($TD_1, TD_2, ..., TD_p$).
We use K-means to partition the dataset since K-means maximizes Euclidean distance between any two clusters.
The samples with short Euclidean distance will be clustered into one group.
After K-means, each subset will have its data center ($CT_1, CT_2, ..., CT_p$).
Then we launch $p$ independent KRRs ($KRR_1, KRR_2, ..., KRR_p$) to process these $p$ subsets.

\begin{figure}[!t]
\centering
\includegraphics[width=2.1in]{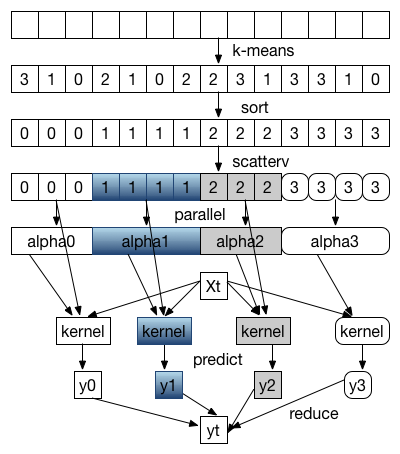}
\caption{Implementation of Kmeans KRR (KKRR). Both k-means and sort can be parallelized. We use the standard MPI implementation for scatter operation.}
\label{fig:kkrr}
\end{figure}

\begin{figure}[ht]
\centering
\renewcommand{\thesubfigure}{\thefigure.\arabic{subfigure}}
\subfigure[]{\includegraphics[width=2.8in]{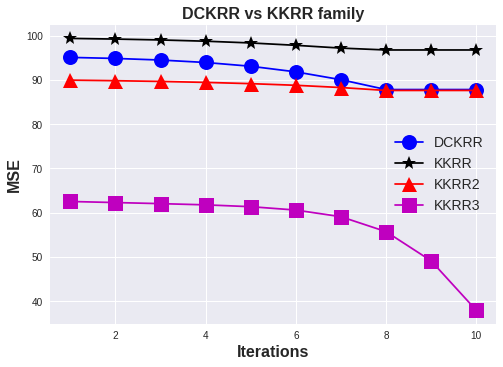}
\label{fig_first_case}}
\subfigure[]{\includegraphics[width=2.8in]{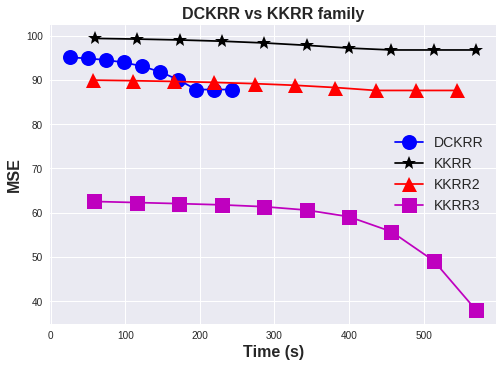}
\label{fig_first_case}}
\caption{Comparison between DC-KRR and KKRR family, using same parameter set on 96 NERSC Edison processors.  The test data set MSD is described in Section \ref{sec:dataset}.  KKRR2 is an accurate but slow algorithm.  KKRR3 is an optimal but unrealistic method for comparison purposes. 
}

\label{fig:dckrr_kkrr_comp}
\end{figure}

\iffalse
\begin{figure*}[ht]
\centering
\renewcommand{\thesubfigure}{\thefigure.\arabic{subfigure}}
\subfigure[]{\includegraphics[width=1.8in]{pic/dckrr_kkrr_iter.png}
\label{fig_first_case}}
\subfigure[]{\includegraphics[width=1.8in]{pic/dckrr_kkrr2_iter.png}
\label{fig_first_case}}
\subfigure[]{\includegraphics[width=1.8in]{pic/dckrr_kkrr3_iter.png}
\label{fig_second_case}}
\subfigure[]{\includegraphics[width=1.8in]{pic/dckrr_kkrr_time.png}
\label{fig_third_case}}
\subfigure[]{\includegraphics[width=1.8in]{pic/dckrr_kkrr2_time.png}
\label{fig_fourth_case}}
\subfigure[]{\includegraphics[width=1.8in]{pic/dckrr_kkrr3_time.png}
\label{fig_fifth_case}}
\caption{The comparison between DC-KRR with KKRR family. They use the same parameter set on 96 NERSC Edison processors. The information of test dataset is in Section \ref{sec:dataset}. Eventually, KKRR2 will achieve a lower MSE than DCKRR. However, because DCKRR is much faster than KKRR2, the slope of DCKRR time-MSE curve is much sharper than that of KKRR2. The definition of KKRR3 is in Section \ref{sec:bkrr_kkrr3}.}
\label{fig:dckrr_kkrr_comp}
\end{figure*}
\fi

Given a test sample $\hat{x}$, instead of only using one model to predict $\hat{x}$, we make all the nodes have a copy of $\hat{x}$.
This additional cost is trivial for two reasons: (1) the test dataset is much smaller than the training dataset, and the training dataset is much smaller than the kernel matrix, which is the major memory overhead.
(2) This broadcast is finished at initial step and there is no need to do it at every iteration.
Each node $i$ makes a prediction $\bar{y}_i$ for $\hat{x}$ by using its local model.
Then we get the final prediction $\tilde{y}$ by a global reduction (average) over all nodes ($\tilde{y} = \sum_{i=1}^p \bar{y}_i/p$).
Figure~\ref{fig:kkrr} is the framework of KKRR. KKRR is highly parallel because all the subproblems are independent of each other. However, the performance of KKRR is not good as we expect. 
From Figure \ref{fig:dckrr_kkrr_comp}.1, we observe that DC-KRR achieves better accuracy than KKRR. In addition, KKRR is slower than DC-KRR.
In the following sections, we will make the algorithm more accurate and scalable, which is better than DC-KRR.  
%We will firstly make the algorithm more accurate and then make it 
%fast and scalable. 

\subsection{KKRR2}
The low accuracy of KKRR is mainly due to its conquer step. 
%We think the reason that KKRR has a lower accuracy is that the conquer step is not effective. 
For DC-KRR, because the divide step is a random and even partition, the 
sub-datasets and local models are similar to each other, and thus averaging works pretty well. 
%The local models generated by sub-datasets are also similar to each other. 
For KKRR, the clustering method divides the original dataset into different sub-datasets which are far away from each other in Euclidean distance. 
%, and these sub-datasets are different from each
%other. 
The local models generated by sub-datasets are also totally different from each other. 
Thus, using their average will get worse accuracy because some models are unrelated to the test sample $\hat{x}$. For example,
if the data center of $i$-th partition ($CT_i$) is far away from $\hat{x}$ in 
Euclidean distance, the $i$-th model should not be used to make prediction for 
$\hat{x}$. Since we divide the original dataset based on Euclidean distance, 
the similar procedure should be used in the prediction phase. 
Thus, we design the following algorithm.

After the training process, each sub-KRR will generate its own model file ($MF_1, MF_2, ..., MF_P$).
We can use these models independently for prediction. For a given unknown sample $\hat{x}$,
if its closest data center (in Euclidean distance) is $CT_i$, we only use $MF_i$ to make a prediction for $\hat{x}$. 
We call this version KKRR2. From Figures \ref{fig:dckrr_kkrr_comp}.1 and \ref{fig:dckrr_kkrr_comp}.2 we observe that KKRR2 is more accurate than DCKRR. However, KKRR2 is slower than DCKRR. For example, to get the same accuracy (MSE=88), KKRR2 is 2.2$\times$ (436s vs 195s) slower than DCKRR.
Thus we need to focus on speed and scalability.

\subsection{Balanced KRR (BKRR)}
\begin{figure}[ht]
%\begin{figure*}[ht]
\centering
\renewcommand{\thesubfigure}{\thefigure.\arabic{subfigure}}
\subfigure[Load Balance for Data Size]{\includegraphics[width=2.7in]{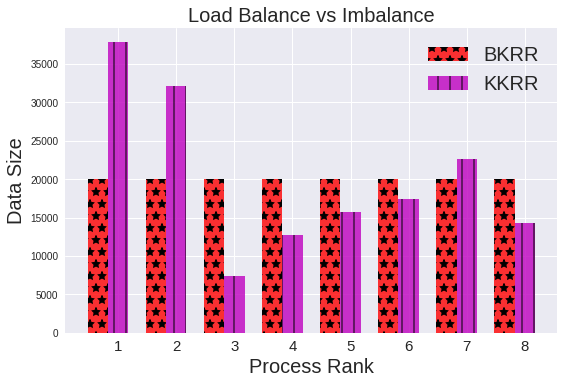}
\label{fig:load_balance1}}
\subfigure[Load Balance for Time]{\includegraphics[width=2.7in]{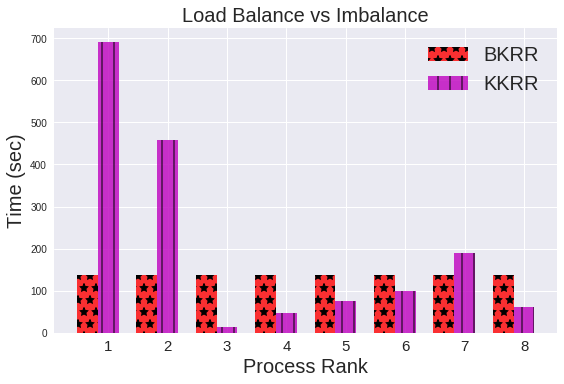}
\label{fig:load_balance2}}
\caption{This experiment is conducted on NERSC Cori Supercomputer \cite{wright2015cori}. We use 8 nodes for load balance test. The test dataset is MSD dataset and we use 16000 samples. We observe that BKRR has roughly 2000 samples on each node, which leads to a perfect load balance. For KKRR, because different nodes have different number of samples, the fastest node is 51$\times$ faster than the slowest node, which leads to a huge resource waste.}
\label{fig:load_balance}
%\end{figure*}
\end{figure}

Based on the profiling results in Figure~\ref{fig:load_balance}, we observe that the major source of KKRR's
poor efficiency is load imbalance. The reason is that the partitioning by K-means is irregular and imbalanced.
% For example, we want a 15-15-15-15 partition if we have 60 samples while K-means may give us a 36-10-11-3 partition.
For example, processor~2 in Figure~\ref{fig:load_balance} has to handle $n$ = 35,137 samples while processor~3
only needs to process $n$ = 7,349 samples. Since the memory requirement grows like $\Theta(n^2)$ and
the number of flops grows like $\Theta(n^3)$, processor~3 runs 51$\times$ faster than processor~1
(Figure~\ref{fig:load_balance}). %This load imbalance issue leads to the poor efficiency of KKRR family.
On the other hand, partitioning by K-means is data-dependent and the sizes of 
clusters cannot be controlled. This makes it unreliable to be used in 
practice, and thus we need to replace K-means with
a load-balanced partitioning algorithm. To this end, we design a K-balance clustering algorithm and use it to build Balanced Kernel Ridge Regression (BKRR).

\begin{algorithm}
\small
\DontPrintSemicolon % Some LaTeX compilers require you to use \dontprintsemicolon instead
\KwIn{\newline$CT$[i] is the center of i-$th$ cluster;\newline$CS$[i] is the size of i-$th$ cluster;\newline$SA$[i] is the i-$th$ sample;\newline$n$ is the number of samples;\newline$P$ is the number of clusters (\#machines);}
\KwOut{\newline$MB$[i] is the closest center to i-$th$ sample;\newline$CT$[i] is the center of i-$th$ cluster;}
Run K-means to get $CT$[$1$], $CT$[$2$], ..., $CT$[$p$]\;
%Randomly pick $p$ samples from $n$ samples ($RS$[0:p])\;
%\For{$i \in 1:p$} {
%    $CT$[$i$] = $RS$[$i$]\;
%}
$balanced$ = $n/p$\;
\For{$i \in 1:n$} {
    $mindis$ = inf\;
    $minind$ = 0\;
    \For{$j \in 1:p$} {
        $dist$ = EuclidDistance($SA$[$i$], $CT$[$j$])\;
        \If{dist $<$ mindis {\bf and} $CS$[$j$] $<$ balanced}{
            $mindis = dist$\;
            $minind = j$\;
        }
    }
    $CS$[$minind$]++\;
	$MB$[$i$] = $minind$\;
}
\For{$i \in 1:p$} {
    $CT$[$i$] = 0\;
}
\For{$i \in 1:n$} {
    $j$ = $MB$[$i$]\;
    $CT$[$j$] += $SA$[$i$]\;
}
\For{$i \in 1:p$} {
    $CT$[$i$] = $CT$[$i$] / $CS$[$i$]\;
}
\caption{K-balance Clustering}
\label{algo:kbc}
\end{algorithm}

\begin{algorithm}
\small
\DontPrintSemicolon % Some LaTeX compilers require you to use \dontprintsemicolon instead
\KwIn{\newline$n$ labeled data points $(x_i, y_i)$ for training; \newline
$k$ labeled data points $(\hat{x}_j, \hat{y}_j)$ for testing;\newline
both $x_i$ and $\hat{x}_j$ are $d$ dimensional vectors;\newline
$i \in \{1, 2, ..., n\}$, $j \in \{1, 2, ..., k\}$;\newline
$\widehat{MSE}$ $\leftarrow$ $\infty$ (Initial Mean Squared Error);}
%$t$ is the rank of a machine, $t \in \{1, ..., p\}$;\newline
\KwOut{\newline Mean Squared Error ($\widehat{MSE}$) of prediction \newline best parameters $\hat{\lambda}, \hat{\sigma}$}
%Parallel IO: read $n/p$ data points on each node\;
$t$ $\leftarrow$ rank of a machine, $t \in \{1, ..., p\}$\;
Do K-balance clustering (Algorithm \ref{algo:kbc})\;
$M$ = [$X$, $y$] is a $(n/p)$-by-$(d+1)$ matrix\;
Store data points $(x_i, y_i)$ $i \in \{1, ..., n/p\}$ as [$X$, $y$]\;
%\For{$i \in 1:p$} {
%	gather $i$-th cluster data on each node to node-$i$
%}
%Copy the gathered data to $M$\;
Create a $m$-by-$m$ kernel matrix $K$, $m = n/p$\newline
\For{$i \in 1:m$} {
	$x_i, y_i$ = M[i][1:d+1]\;
	Machine $t$: $(x_i^t, y_i^t)$ = $(x_i, y_i)$, $t \in \{1, ..., p\}$\;
}
\For{$\lambda \in \Lambda$ and $\sigma \in \Sigma$} {
\For{$i \in 1:m$} {
    \For{$j \in 1:m$} {
        $K[i][j]$ = $\Phi$($x_i$, $x_j$) based on Table \ref{table:kernels}\;
    }
}
Solve linear equation $(K+\lambda m I) \alpha = y$ for $\alpha$\;
\For{$t \in 1:p$} {
	$err_t \leftarrow 0$
}
\For{$j \in 1:k$} {
    \If{$t = MyCluster(\hat{x}_j)$} {
    	$\tilde{y}_j \leftarrow {\sum}_{i=1}^{m} {\alpha}_i^t K(x_i^t, \hat{x}_j)$\;
	$err_t \leftarrow err_t + ||\tilde{y}_j-\hat{y}_j||^2$
    }
}
Reduce: $MSE =  ({\sum}_{t=1}^{p} err_t)/k$\;
\If{$t = 0$} {
	\If{$MSE < \widehat{MSE}$} {
    		$\widehat{MSE} \leftarrow MSE$, $\widehat{\lambda} \leftarrow \lambda$, $\widehat{\sigma} \leftarrow \sigma$\;
	}
}
}
\caption{Balanced KRR2 (BKRR2)}
\label{algo:bkrr2}
\end{algorithm}

In our design, a machine corresponds to a clustering center. If we have $p$ machines,
then we partition the training dataset into $p$ parts. As mentioned above, the objective
of K-balance partitioning algorithm is to make the number of samples on each node close to $n/p$.
If a data center has $n/p$ samples, then we say it is balanced. The basic idea of this algorithm
is to find the closest center for each sample, and if a given data center has been balanced,
no additional sample will be added to this center. The detailed K-balance clustering
method is in Algorithm \ref{algo:kbc}. Line 1 of Algorithm \ref{algo:kbc} is an important step because K-balance needs to
first run K-means algorithm to get the data centers. This makes K-balance have a similar clustering pattern as K-means.
Lines 3-12 find the center for each sample. Lines 6-10 find the best under-load center
for the $i$-th sample. Lines 15-19 recompute the data center by averaging all the samples
in a certain center. Recomputing the centers by averaging is optional because it will not
necessarily make the results better. From Figure~\ref{fig:load_balance} we observe that K-balance
partitions the dataset in a balanced way. After K-balance clustering, all the nodes have
the same number of samples, so in the training phase all the nodes roughly 
have the same training time and memory requirement.
%K-balance reserves the idea of grouping the samples based on Euclidean distance.

After replacing K-means with K-balance, KKRR becomes BKRR, and KKRR2 becomes BKRR2. Algorithm \ref{algo:bkrr2} is a framework of BKRR2.
As we mentioned in the sections of Abstract and Introduction, KKRR2 is the optimized version of KKRR and BKRR2 is the optimized version of BKRR.
Lines 1-7 perform the partition. Lines 9-22 perform one iteration of the training.
Lines 9-11 construct the kernel matrix. Line 12 solves the linear equation on each machine. Lines 13-18 perform the prediction. Only the best model does the prediction for a certain test sample.
Lines 19-22 evaluate the accuracy of the system.
%We remove all the communication costs of BKRR2 in the training part. 
There is no communication cost of BKRR2 during training except for the initial data partition and the final model selection.
The additional cost of BKRR2 comes from two parts: (1) the K-balance algorithm and partition, and (2) the prediction.

The overhead of K-balance is tiny compared with the training part. K-balance first does K-means. The cost of K-means is $\Theta(In)$ where
$I$ is the number of K-means iterations, which is usually less 100. The cost of lines 3-12 in Algorithm \ref{algo:kbc} is $\Theta(pn)$ where $p$ is the number of
partitions and also the number of machines. $\Theta(In+pn)$ is tiny compared with the training part, which is $\Theta(|\Sigma||\Lambda| n^3/p^3)$.
%Thus, we do not even need to parallelize K-balance in this situation, which leads to no communication.
For example, if we use BKRR2 to process the 32k MSD training samples on 96 NERSC Edison processors, the single-iteration training time is 20 times larger than the K-balance and partition time.
Since the computational overhead of K-balance is low compared to KRR training ($n = \Theta(p^4)$ in practice), we can just use one node to finish the K-balance algorithm.
Although it is not necessary, we have an approximate parallel algorithm for K-balance. We conduct parallel K-means clustering and distribute the centers to all the nodes. Then we partition the samples to all the processors and set $balanced$ as $n/p^2$. Since $n = \Theta(p^4)$ in practice, this parallel implementation roughly gets the same results with the single-node implementation. The overhead of this approximate parallel algorithm is $\Theta(In/p+n)$.

\begin{figure}[!t]
\centering
\includegraphics[width=3.3in]{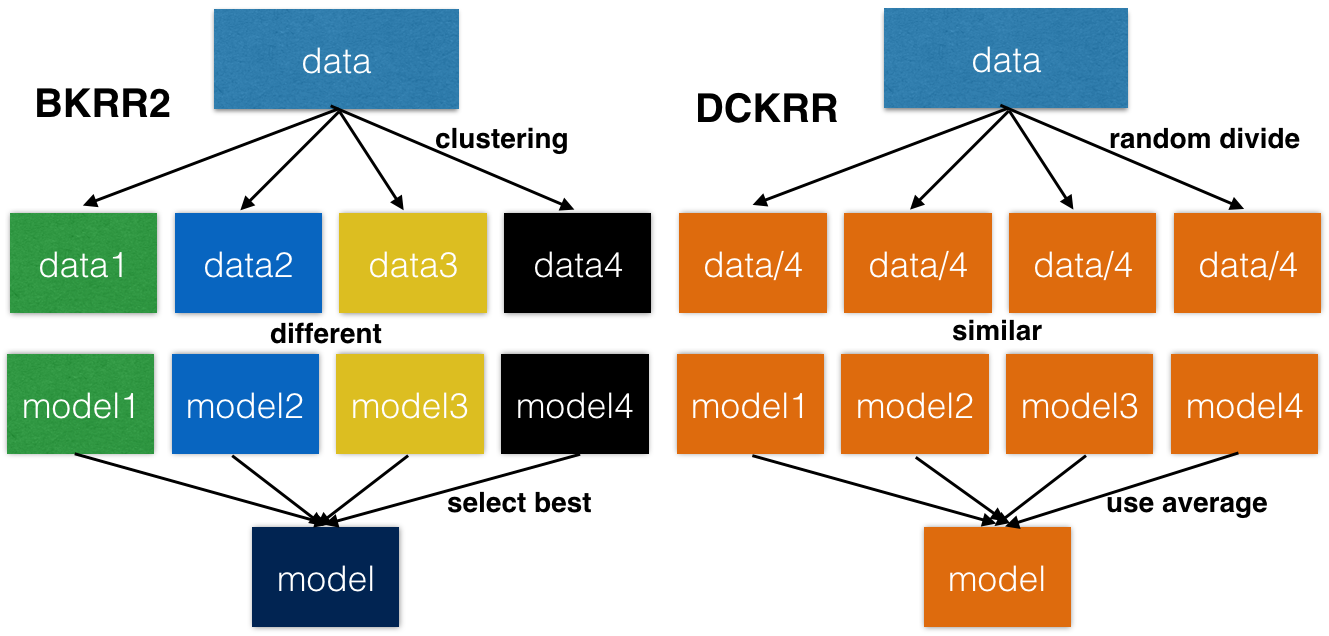}
\caption{Difference between BKRR2 and DCKRR. BKRR2: partition the dataset into $p$ different parts, generate $p$ different models, and select the best model. DCKRR: partition the dataset into $p$ similar parts, generate $p$ similar models, and use the average of all the models.}
\label{fig:cakrr_dckrr_diff}
\end{figure}

For the prediction part, instead of conducting $k$ small communications, we make each machine first compute its own error (line 18 of Algorithm \ref{algo:bkrr2}) and then only conduct one global communication (line 19 of Algorithm \ref{algo:bkrr2}).
The reason is detailed below.
The runtime on a distributed system can be estimated by the sum of the time 
for computation and the time for communication: 
$f \times \gamma + n_m \times (\alpha + n_b \times \beta)$, 
where $f$ is the number of floating point operations,  $\gamma$ is the time per floating point operation, $\alpha$ is latency, $1/\beta$ is the bandwidth, $n_m$ is the number of messages, and $n_b$ is the number of bytes per message. In practice, $\alpha \gg \beta \gg \gamma$ (e.g. $\alpha=7.2\times10^{-6}$s, $\beta=0.9 \times 10^{-9}$s for Intel NetEffect NE020, $\gamma=2 \times 10^{-11}$s for Intel 5300-series Clovertown). Therefore, one big message is much cheaper than $k$ small messages because $k \times (\alpha + n_b \times \beta) \gg (\alpha + k \times n_b \times \beta)$. This optimization reduces the latency overhead. 
Figures \ref{fig:dckrr_cakrr_comp} and \ref{fig:dckrr_cakrr_example} show that BKRR2 achieves lower error rate than DC-KRR in a shorter time for a variety of datasets.
Figure \ref{fig:cakrr_dckrr_diff} shows the difference between BKRR2 and DC-KRR.
%\textcolor{blue}{**************************************** I stopped here ****************************************}

\subsection{BKRR3 and KKRR3: Error Lower Bound and The Unrealistic Approach\label{sec:bkrr_kkrr3}}
%As we mentioned before, the idea of KKRR and BKRR family is to partition the data into $p$ parts and make $p$ different models. Then we select the best model from these $p$ models for each test sample.
The KKRR and BKRR families share the same idea of partitioning the data into $p$ parts and making $p$ different models, but they are different in that the KKRR family is optimized for accuracy while the BKRR family is optimized for performance.
We want to know the gap between our method and the best theoretical method.
Let us refer to the theoretical KKRR method as KKRR3 and the theoretical BKRR method as BKRR3.

\begin{figure*}[ht]
\centering
\renewcommand{\thesubfigure}{\thefigure.\arabic{subfigure}}
\subfigure[1024 training, 361 test samples]{\includegraphics[width=1.8in]{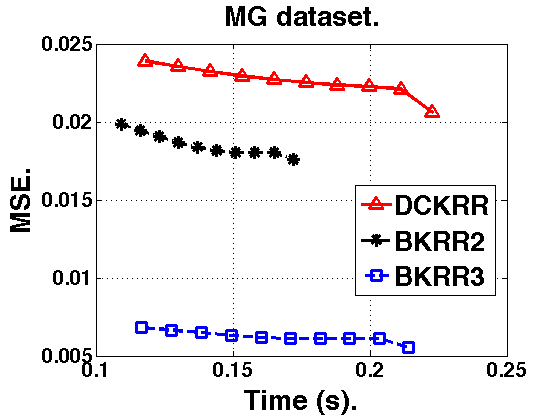}
\label{fig_first_case}}
\subfigure[2560 training, 547 test samples]{\includegraphics[width=1.8in]{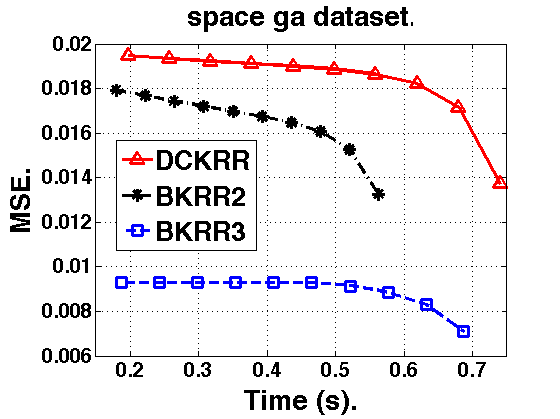}
\label{fig_first_case}}
%\subfigure[7680 train samples, 512 test samples]{\includegraphics[width=2.2in]{pic/cpusmall.png}
%\label{fig_second_case}}
%\subfigure[4096 train samples, 1024 test samples]{\includegraphics[width=2.2in]{pic/e2006.png}
%\label{fig_third_case}}
\subfigure[18432 training, 2208 test samples]{\includegraphics[width=1.8in]{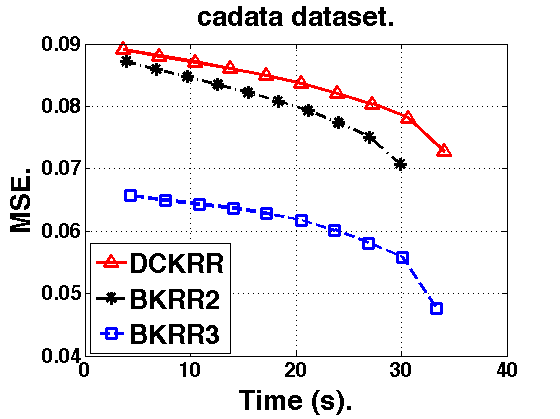}
\label{fig_fourth_case}}
%\subfigure[4096 train samples, 81 test samples]{\includegraphics[width=2.2in]{pic/abalone.png}
%\label{fig_fifth_case}}
\caption{These figures do a point-to-point comparison between BKRR2, BKRR3 and DC-KRR. They use the same parameter set and conduct the same number of iterations on 96 NERSC Edison processors. BKRR2 is faster than DCKRR for getting the same accuracy. BKRR3 is an optimal but unrealistic implementation for comparison purposes.}
\label{fig:dckrr_cakrr_example}
\end{figure*}

%Let us detail BKRR3 here. 
BKRR3 is similar to BKRR2 in terms of communication and computation pattern. Like BKRR and BKRR2, after the training process, each sub-BKRR will generate its own model file ($MF_1, MF_2, ..., MF_P$).
We can use these model files independently for prediction. For a given test sample $\hat{x}$ ($\hat{y}$ is its true regressand), we make all the nodes have a copy of $\hat{x}$ (like KKRR).
Each model will make a prediction for $\hat{x}$. We get $\tilde{y}_1$, $\tilde{y}_2$, ..., $\tilde{y}_p$ from $p$ models respectively.
We select $MF_i$ for prediction where $i = {argmin}_j||\tilde{y}_j-\hat{y}||^2$. 
This means we {\bf inspect} the true regressand to make sure we select the best model for each test sample.
When changing K-balance to K-means, BKRR3 becomes KKRR3,
which is much more accurate than DCKRR (Figures \ref{fig:dckrr_kkrr_comp}.1 and \ref{fig:dckrr_kkrr_comp}.2).
The MSE of BKRR3 is the lower bound of the MSE of BKRR2 because BKRR3 can always use the best model for each test sample. 
%We refer to BKRR3 as CA-KRR (Communication-Avoiding KRR) because it is the ideal and unrealistic version of our method.
In fact, BKRR3 is even much more accurate than the original method (DKRR) for all the testings in our experiments. The framework of BKRR3 is in Algorithm \ref{algo:bkrr3}.
Figures \ref{fig:dckrr_cakrr_comp} and \ref{fig:dckrr_cakrr_example} show the results.
BKRR3 is always the best approach in these comparisons. BKRR3 achieves much higher accuracy than DCKRR and BKRR2.

%From Fig. \ref{fig:dckrr_cakrr_comp} we can observe that BKRR, BKRR2, and CA-KRR have the same speed with DC-KRR.
%BKRR is not accurate than DCKRR, BKRR2 is accurate than DCKRR, and CA-KRR is much more accurate than DC-KRR.
%The reason is the same as we explain in the section of KKRR. 
%Fig. \ref{fig:dckrr_cakrr_example} shows that CA-KRR achieves lower error rate than DC-KRR in a variety of datasets.
%The reason why we refer to BKRR3 as CA-KRR is that only the data partition and prediction parts have the communication costs. The training part (the major overhead) is totally embarrassingly parallel, which has no communication. 

%We formulate it here, $t_1$ is the partition time and $t_2$ is the training time. $t_2$ is much larger than $t_1$. For example $t_2 = 20t_1$. The total time is $|\Sigma||\Lambda| t_2 + t_1$. $|\Sigma|$ and $|\Lambda|$ are usually more than 30, so $|\Sigma||\Lambda|$ is more than 1000. Thus $t_1$ is very tiny compared with $t_1$+1000$t_2$.
%\textcolor{blue}{The communication cost between CA-KRR and D-KRR: the reason why we call this method as CA-KRR. counting the Gflops and write out the communication equations by $n$ and $p$, get some data to count the real communication cost. }

\begin{figure}[ht]
\centering
\renewcommand{\thesubfigure}{\thefigure.\arabic{subfigure}}
\subfigure[]{\includegraphics[width=2.8in]{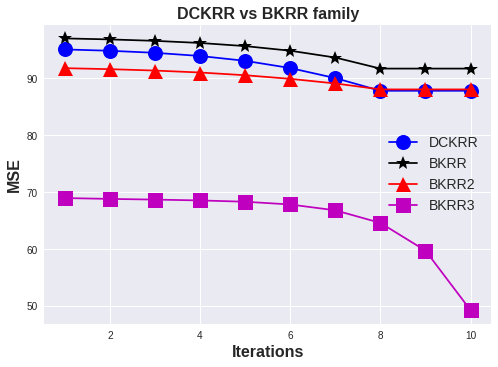}
\label{fig_first_case}}
\subfigure[]{\includegraphics[width=2.8in]{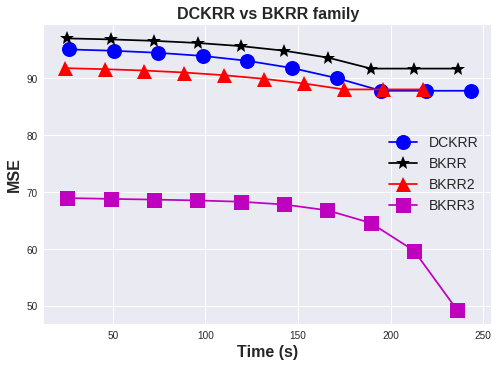}
\label{fig_first_case}}
\caption{These figures do a point-do-point comparison between DC-KRR and BKRR family using test data set MSD (described in Section \ref{sec:dataset}). They use the same parameter set and conduct the same number of iterations on 96 NERSC Edison processors. DCKRR is more accurate than BKRR. BKRR2 is faster than DCKRR for getting the same accuracy. BKRR3 is an optimal but unrealistic implementation for comparison purposes.}

\label{fig:dckrr_cakrr_comp}
\end{figure}

\iffalse
\begin{figure*}[ht]
\centering
\renewcommand{\thesubfigure}{\thefigure.\arabic{subfigure}}
\subfigure[]{\includegraphics[width=1.8in]{pic/dckrr_bkrr1_iter.png}
\label{fig_first_case}}
\subfigure[]{\includegraphics[width=1.8in]{pic/dckrr_bkrr2_iter.png}
\label{fig_first_case}}
\subfigure[]{\includegraphics[width=1.8in]{pic/dckrr_bkrr3_iter.png}
\label{fig_second_case}}
\subfigure[]{\includegraphics[width=1.8in]{pic/dckrr_bkrr1_time.png}
\label{fig_third_case}}
\subfigure[]{\includegraphics[width=1.8in]{pic/dckrr_bkrr2_time.png}
\label{fig_fourth_case}}
\subfigure[]{\includegraphics[width=1.8in]{pic/dckrr_bkrr3_time.png}
\label{fig_fifth_case}}
\caption{These 6 figures show a point-to-point comparison between DC-KRR with BKRR family. They use the same parameter set and conduct the same number of iterations on 96 NERSC Edison processors.}
\label{fig:dckrr_cakrr_comp}
\end{figure*}
\fi

\begin{figure}[!t]
\centering
\includegraphics[width=2.7in]{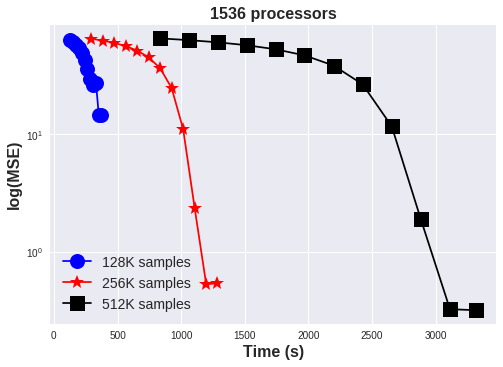}
\caption{BKRR2 results based on MSD dataset. Using increasing number of samples on a fixed number of processors, we can get a better model but also observe the time increase in the speed of $\Theta(n^3)$.}
\label{fig:accuracy_scaling}
\end{figure}

\begin{figure*}[ht]
\centering
\renewcommand{\thesubfigure}{\thefigure.\arabic{subfigure}}
\subfigure[Time]{\includegraphics[width=1.8in]{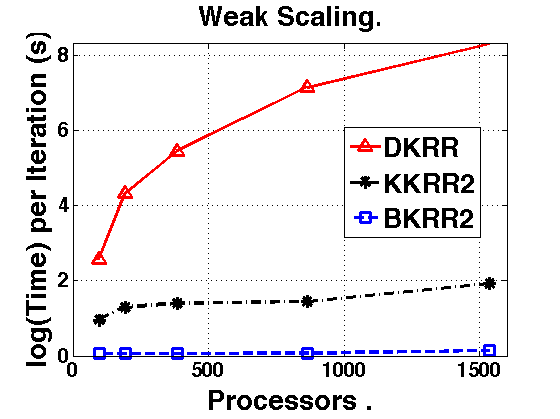}
\label{fig:weak_scaling1}}
\subfigure[Efficiency]{\includegraphics[width=1.8in]{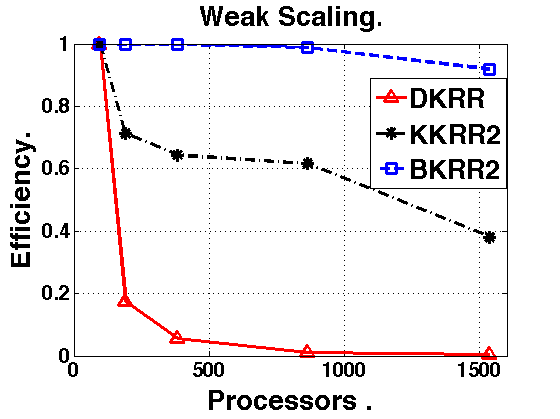}
\label{fig:weak_scaling2}}
\subfigure[Accuracy]{\includegraphics[width=1.8in]{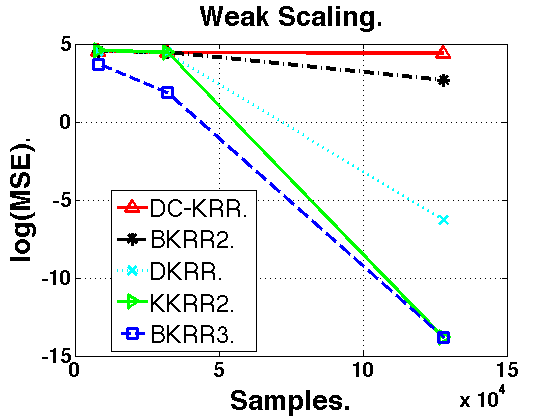}
\label{fig:weak_scaling3}}
\caption{Results based on MSD dataset. We use 96 processors (i.e. 4 nodes) and 8k samples as the baseline. The lowest MSE of DKRR is 0.001848 on 1536 processors. The lowest MSE of BKRR3 is $10^{-7}$. The weak scaling efficiency of DKRR at 1536 processors is 0.32\%. The weak scaling efficiency of KKRR2 and BKRR2 at 1536 processors is 38\% and 92\%, respectively. The MSE of DC-KRR for 2k test samples only decreases from 88.9 to 81.0, which is a bad weak scaling accuracy. The MSE of BKRR2 decreases from 93.1 to 14.7. The MSE of KKRR2 decreases from 95.0 to $10^{-7}$. The data is in Tables \ref{table:weak_scaling} and \ref{table:accuracy_scaling}. We conclude that the proposed methods outperform the existing methods.}
\label{fig:weak_scaling_comp}
\end{figure*}

\begin{algorithm}
\small
\DontPrintSemicolon % Some LaTeX compilers require you to use \dontprintsemicolon instead
\KwIn{\newline$n$ labeled data points $(x_i, y_i)$ for training; \newline
another $k$ labeled data points $(\hat{x}_j, \hat{y}_j)$ for testing;\newline
both $x_i$ and $\hat{x}_j$ are $d$ dimensional vectors;\newline
$i \in \{1, 2, ..., n\}$, $j \in \{1, 2, ..., k\}$;\newline
$t$ is the rank of a machine, $t \in \{1, ..., p\}$;\newline}
\KwOut{\newline Mean Squared Error ($\widehat{MSE}$) of prediction \newline best parameters $\hat{\lambda}, \hat{\sigma}$}
%Parallel IO: read $n/p$ data points on each node\;
Do K-balance clustering\;
$M$ = [$X$, $y$] is a $(n/p)$-by-$(d+1)$ matrix\;
Store data points $(x_i, y_i)$ $i \in \{1, ..., n/p\}$ as [$X$, $y$]\;
%\For{$i \in 1:p$} {
%	gather $i$-th cluster data on each node to node-$i$
%}
%Copy the gathered data to $M$\;
Create a $m$-by-$m$ kernel matrix $K$, $m = n/p$\newline
\For{$i \in 1:m$} {
	$x_i, y_i$ = M[i][1:d+1]\;
	Machine $t$: $(x_i^t, y_i^t)$ = $(x_i, y_i)$, $t \in \{1, ..., p\}$\;
}
\For{$\lambda \in \Lambda$ and $\sigma \in \Sigma$} {
\For{$i \in 1:m$} {
    \For{$j \in 1:m$} {
        $K[i][j]$ = $\Phi$($x_i$, $x_j$) based on Table \ref{table:kernels}\;
    }
}
Solve linear equation $(K+\lambda m I) \alpha = y$ for $\alpha$\;
\For{$j \in 1:k$} {
    Machine $t$: $\tilde{y}_j^t = {\sum}_{i=1}^{m} {\alpha}_i^t K(x_i^t, \hat{x}_j)$\;
    Global reduce: $id = {argmin}_t||\tilde{y}_j^t-\hat{y}_j||^2$\;
    Send/Receive: $\tilde{y}_j = \tilde{y}_j^{id}$\;
}
\If{$rank = 0$} {
	$MSE = \frac{1}{k} {\sum}_{j=1}^{k} (\tilde{y}_j  - \hat{y}_j)^2$\;
	\If{$MSE < \widehat{MSE}$} {
    		$\widehat{MSE} \leftarrow MSE$, $\widehat{\lambda} \leftarrow \lambda$, $\widehat{\sigma} \leftarrow \sigma$\;
	}
}
}
\caption{BKRR3}
\label{algo:bkrr3}
\end{algorithm}

\section{Implementation and Analysis}
\subsection{Real-World Dataset}\label{sec:dataset}
To give a fair comparison with DC-KRR, we use the Million Song Dataset (MSD) \cite{bertin2011million} as our major dataset in our experiments because MSD was used in the paper of DC-KRR. It is a freely-available collection of audio features and metadata for a million contemporary popular music tracks.
The dataset contains $n=$ 515,345 samples. Each sample is a song (track) released between 1922 and 2011, and the song is represented as a vector of timbre information computed from the song. Each sample consists of a pair of $(x_i, y_i)$ where $x_i$ is a $d$-dimensional ($d=90$) vector and $y_i \in [1922, 2011]$ is the year that the song was released.
The Million Song Dataset Challenge aims at being the best possible offline evaluation of a music recommendation system.
It is a large-scale, personalized music recommendation challenge, where the goal is to predict the songs that a user will listen to, given both the user's listening history and full information (including meta-data and content analysis) for all songs. 
To justify the efficiency of our approach, we use another three real-world datasets. The information of these four datasets are summarized in Table \ref{table:dataset}. All these datasets were downloaded from \cite{regressiondata}.

\begin{table}[ht]
\footnotesize
\renewcommand{\arraystretch}{1.3}
\caption{The test datasets}
\centering
\begin{tabular}{l*{9}{c}r}
\hline
{\bf name} & MSD & cadata  & MG & space-ga\\
\hline
{\bf \# Train} & 463,715 & 18,432 & 1,024 & 2,560\\
\hline
{\bf \# Test} & 51,630 & 2,208 & 361 & 547\\
\hline
{\bf Dimension} & 90 & 8 & 6 & 6\\
\hline
{\bf Application} & Music  & Housing & Dynamics & Politics\\
\hline
\end{tabular}
\label{table:dataset}
\end{table}

\begin{table}[ht]
\footnotesize
\renewcommand{\arraystretch}{1.3}
\caption{Weak Scaling in time. We use 96 processors and 8000 MSD samples as the baseline. Constant time means perfect scaling. BKRR2 has very good scaling efficiency. DKRR's scaling efficiency is poor.}
\centering
\begin{tabular}{l*{12}{c}r}
\hline
Method & Processors & 96 & 192 & 384 & 768 & 1536\\
\hline
BKRR2 & IterTime (s) & 1.06 & 1.06 & 1.06 & 1.08 & 1.15\\
\hline
KKRR2 & IterTime (s) & 2.60 & 3.66 & 4.05 & 4.23 & 6.85\\
\hline
DKRR & IterTime (s) & 13 & 75.1 & 234 & 1273 & 4048\\
\hline
BKRR2 & Efficiency & 1.0 & 1.0 & 1.0 & 0.99 & 0.92\\
\hline
KKRR2 & Efficiency & 1.0 & 0.71 & 0.64 & 0.62 & 0.38\\
\hline
DKRR & Efficiency & 1.0 & 0.17 & 0.06 & 0.01 & 0.003\\
\hline
\end{tabular}
\label{table:weak_scaling}
\end{table}

\begin{table}[ht]
\footnotesize
\renewcommand{\arraystretch}{1.3}
\caption{Weak Scaling in Accuracy on MSD Dataset. Lower is better. DCKRR is a bad algorithm.}
\centering
\begin{tabular}{l*{9}{c}r}
\hline
Samples & DCKRR & BKRR2  & DKRR & KKRR2 & BKRR3\\
\hline
8k & 88.9 & 93.1 & 90.9 & 95.0 & 40.2\\
\hline
32k & 85.5 & 87.7 & 85.0 & 87.5 & 6.6\\
\hline
128k & 81.0 & 14.7 & 0.002 & $10^{-7}$ & $10^{-7}$\\
\hline
\end{tabular}
\label{table:accuracy_scaling}
\end{table}

\subsection{Fair Comparison}
Let us use $p$ as the number of partitions or nodes, $\rho$ as the number of processors.
Each node has 24 processors. When we use $\rho$=1536 processors, we actually divide the dataset into $p$=64 parts.
Each machine generates a local model for BKRR2. 
To give a fair comparison, we make sure all the comparisons were tuned based on the same parameter set. 
Different methods may select different best-parameters from the same parameter set to achieve its lowest MSE.
%To avoid confusing the readers that we may select some parameters that favor our methods, our source code is available online\footnote{\url{https://people.eecs.berkeley.edu/~youyang/cakrr.zip}}. 
From Figures~\ref{fig:dckrr_cakrr_comp},~\ref{fig:dckrr_cakrr_example} and \ref{fig:weak_scaling_comp}, 
we clearly observe BKRR2 is faster than DC-KRR and also achieves lower prediction error on all the datasets. 
In other words, BKRR2 and DC-KRR may use different parameters to achieve their lowest MSEs. The lowest MSE of BKRR2 is lower than the lowest MSE of DC-KRR.
On the other hand, BKRR2 is slightly faster than both DC-KRR and BKRR3 (for both single iteration time and overall time). The reason is that each machine only needs to process $1/p$ of the test samples for prediction. For DC-KRR and BKRR3, each machine needs to process all the test samples for prediction.
BKRR2 achieves 1.2$\times$ speedup over DC-KRR on average and has a much lower error rate for 128k-sample MSD dataset (14.7 vs 81.0).
%To the best of our knowledge, 
%Based on our numerous experimental results, the lowest MSE of CA-KRR is lower than that of DC-KRR and DKRR, for each dataset.
%0.740192/0.562867+34.039447/29.868731+193.359865/174.748276+0.222948/0.171949

Because of DKRR's poor weak scaling, BKRR2 runs much faster than DKRR for 1536 processors and 128k training samples. The single iteration time of BKRR2, $t_b$, is 1.15 sec while the single iteration time of DKRR, $t_d$, is 4048 sec.
Here, single iteration means picking a pair of parameters and solving the linear equation once (e.g. lines 9-22 of Algorithm \ref{algo:bkrr2} are one iteration).
The algorithm can get a better pair of parameters after each iteration.
All the algorithms in this paper run the same number of iterations because they use the same parameter tuning space.
However, it is unfair to say BKRR2 achieves 3505$\times$ speedup over DKRR because, for the same 2k test dataset, the lowest MSE of BKRR2 is 14.7 while that of DKRR is 0.002. 
This means the BKRR2 model with 128K samples model ($bm_{128}$) is worse than DKRR model ($dm_{128}$) for accuracy.
To give a fair comparison, we increase the training samples of BKRR2 to 256k so that its lowest MSE can generate a better model ($bm_{256}$). By using $bm_{256}$, the lowest MSE of BKRR2 is 0.53 (Figure \ref{fig:accuracy_scaling}). 
We run it on the same 1536 processors and observe $t_b$ becomes 5.6 sec. In this way, we can say that BKRR2 achieves 723$\times$ speedup over DKRR for achieving roughly the same accuracy by using the same hardware and the same test dataset.
It is worth noting that the biggest dataset that DKRR can handle on 1536 processors is 128k samples, otherwise it will collapse. Therefore, $dm_{128}$ is the best model DKRR can get on 1536 processors.
However, BKRR2 can use a bigger dataset to get a better model than $bm_{256}$ on 1536 processors (e.g. 512k model in Figure \ref{fig:accuracy_scaling}). 

The theoretical speedup of $bm_{128}$ over $dm_{128}$ is the ratio of $\Theta(n^3/p)$ and $\Theta((n/p)^3)$, which is 4096$\times$ for $p$ = 64 and $\rho$ = 1536. We achieve 3505$\times$ speedup.
The theoretical speedup of $bm_{256}$ over $dm_{128}$ is the ratio of $\Theta(n^3/p)$ and $\Theta((2n/p)^3)$, which is 512$\times$ for $p$ = 64 and $\rho$ = 1536. We achieve 723$\times$ speedup.
The difference between theoretical speedup and practical speedup comes from systems and low-level libraries (e.g. the implementation of LAPACK and ScaLAPACK).
%If the readers still think the comparison between BKRR2 and DKRR is unfair because BKRR2 used more data. 
In this comparison, BKRR2's better scalability allows it to use more data than DKRR, which cannot run on an input of the same size. We also want to compare KKRR2 to DKRR for using the same amount of data.
Let us refer to the model generated by KKRR2 using 128k samples as $km_{128}$ and the single iteration time as $t_k$.
$t_k$ is 6.9 sec in our experiment. The MSE of $km_{128}$ is $10^{-7}$, which is even lower than the MSE of $dm_{128}$ (0.002). Thus, KKRR2 achieves 591$\times$ speedup over DKRR for the same accuracy by using the same data and hardware (Table \ref{table:weak_scaling} and \ref{table:accuracy_scaling}).

\subsection{Weak-Scaling Issue}
As we mentioned in a previous section, weak scaling means we fix the data size per node and increase the number of nodes. We use from 96 to 1536 processors (4 to 64 partitions) for scaling tests on the NERSC Edison supercomputer.
The test dataset is the MSD dataset. We set the 96-processor case as the baseline and assume
it has a 100\% weak-scaling. We double the number of samples as we double the number of processors.
Figure \ref{fig:weak_scaling_comp} and Table \ref{table:weak_scaling} show the time weak-scaling and accuracy weak-scaling of BKRR.
In terms of time weak-scaling, BKRR2 achieves 92\% efficiency on 1536 processors while DKRR only achieves 0.32\%.
We then compare the weak scaling in terms of test accuracy: the MSE of DC-KRR for 2k test samples only decreases from 88.9 to 81.0, while the MSE of BKRR2 decreases from 93.1 to 14.7.
KKRR2 reduced the MSE from 95 to $10^{-7}$.
%DKRR has a bad time weak-scaling and DC-KRR has a poor accuracy weak-scaling.
%BKRR2 has a good weak scaling behavior both in system and accuracy.
In conclusion, we observe that DKRR has a bad time scaling and DC-KRR
has a poor accuracy weak scaling. 
Our proposed method, BKRR2, has good weak scaling behavior both in terms of time and accuracy.
The weak scaling of BKRR3 can be shown in Table \ref{table:cakrr_weak_scaling}.

\begin{table}[ht]
\footnotesize
\renewcommand{\arraystretch}{1.3}
\caption{Weak scaling results of BKRR3. We use 96 processors and 8k samples as the baseline. We double the number of samples as we double the number of processors.%The number of samples in each case is 2k. The training samples and test samples are different from each other.
}
\centering
\begin{tabular}{l*{9}{c}r}
\hline
{\bf processors} & 96 & 192 & 384 & 768 & 1536\\
\hline
{\bf Time (s)} & 1311 & 1313 & 1328 & 1345 & 1471\\
\hline
{\bf Efficiency} & 100\% & 99.8\% & 98.7\% & 97.5\% & 89.1\%\\
\hline
{\bf MSE} & 40.2 & 16.5 & 6.62 & 1.89 & $10^{-7}$\\
\hline
\end{tabular}
\label{table:cakrr_weak_scaling}
\end{table}

\subsection{Method Selection}
KKRR2 achieves better accuracy than the state-of-the-art method. If the users need the most accurate method, KKRR2 is the best choice. KKRR2 runs slower than DC-KRR but much faster than DKRR.
BKRR2 achieves better accuracy and also runs faster than the DCKRR. 
BKRR2 is much faster than DKRR with the same accuracy. 
The reason why BKRR2 is more efficient than DKRR is not that it assumes more samples, but rather, because it has much better weak scaling, so it is easier to use more samples to get a better model.
Thus, BKRR2 is the most practical method because it balances the speed and the accuracy.
%The detailed efficiency analysis of BKRR2 is at Section \ref{sec:efficiency}.
BKRR3 can be used to evaluate the performance of systems both in accuracy and speed.
KKRR2 is optimized for accuracy.
We recommend using either BKRR2 or KKRR2.
%Because of DKRR's poor weak scaling, CA-KRR achieves 2000$\times$ speedup (Fig. \ref{fig:weak_scaling_comp}) over DKRR for achieving the same MSE by the same data, the parameter set, and the same platform. 
%Compared with DC-KRR, CA-KRR achieves much lower error rate with the same speed (Fig. \ref{fig:dckrr_cakrr_example}) by the same data, the same parameter set, and the same platform.

\subsection{Implementation Details}
We use CSR (Compressed Row Storage) format for processing the sparse matrix (the original $n$-by-$d$ input matrix, not the dense kernel matrix) in our implementation. We use MPI for distributed processing. To give a fair comparison, we use ScaLAPACK \cite{choi1995scalapack} for solving the linear equation on distributed systems and LAPACK \cite{anderson1999lapack} on shared memory systems, both of which are Intel MKL versions. 
%The K-means partitioning in KKRR is a distributed version, which achieved the same partitioning result and comparable performance with Liao's implementation \cite{parallelkmeans}. 
Our experiments are conducted on NERSC Cori and Edison supercomputer systems \cite{nersc2016}.
%Let us use $p$ as the number of partitions or machines, $\rho$ as the number of processors.
%Each Edison machine has 24 processors. When we use $\rho$=1536 processors, we actually divide the dataset into $p$=64 parts.
%Each machine generates a local model for BKRR2. 
Our source code is available online\footnote{\url{https://people.eecs.berkeley.edu/~youyang/cakrr.zip}}.

The kernel matrix is symmetric positive definite (SPD) \cite{wehbe2003}, and so is $K + \lambda n I$.
Thus we use Cholesky decomposition to solve $(K + \lambda n I)\alpha = y$ in DKRR, which is 2.2$\times$ faster than the Gaussian Elimination version.
%Let us refer to the version using Gaussian Elimination as DKRR2.
%We observe DKRR achieves a 2.2$\times$ speedup over DKRR2 for processing 16k samples on 96 NERSC Edison nodes.

To conduct a weak scaling study, we select a subset (e.g. 64k, 128k, 256k) from 463,715 samples as the training data to generate a model. Then we use the model to make a prediction. We use 2k test samples to show the MSE comparisons among different methods. If we change the number of test samples (e.g. change 2k to 20k), the testing accuracy will be roughly the same. The reason is that all the test samples are randomly shuffled so the 2k case will have roughly the sample pattern as the 20k case. 

\section{Related Work}
The most related work to this paper is DC-KRR \cite{zhang2013divide}. The difference between DC-KRR and this work was presented in the Introduction section.
%There are papers on using low rank techniques to approximate the kernel matrix.
Scholkopf et al \cite{scholkopf1998nonlinear} conducted a nonlinear form of
principal component analysis (PCA) in high-dimensional feature spaces.
Fine et al \cite{fine2002efficient} used Incomplete Cholesky Factorization (ICF) to design an efficient Interior Point Method (IPM). 
Williams et al \cite{williams2001using} designed a Nystr{\"o}m sampling method by carrying out an eigendecomposition on a smaller dataset.
Si et al \cite{si2014memory} designed a memory efficient kernel approximation method by first
dividing the kernel matrix and then doing low-rank approximation on the smaller matrices.
These methods can reduce the running time from $\Theta(n^2d)$ to $\Theta(nd^2)$ 
or $\Theta(ndk)$ where $k$ is the rank of the kernel approximation.
However, all of these methods are proposed for serial analysis on single node systems.
On the other hand, Kernel method is more accurate than existing approximate methods \cite{zhang2013divide}. 
Zhang et al. \cite{zhang2013divide} showed DC-KRR can beat all the previous approximate methods. DC-KRR is considered as state-of-the-art approach.
Thus, we focus on the comparison with Zhang et al. \cite{zhang2013divide}.

The second line of work is to use iterative methods such as gradient descent \cite{yao2007early},
block Jacobi method \cite{schreiber86} and conjugate gradient method\cite{blanchard2010optimal} to reduce the running time. 
However, iterative method can not make full use of computational powers efficiently because they only load a small amount of data to memory. If we scale the algorithm to 100+ cores, Kernel matrix method will be much faster.
These methods provide a trade-off between time and accuracy, and we reserve them for future research.
ASKIT \cite{march2015askit} used n-body ideas to reduce the time and storage of kernel matrix evaluation, which is a direction of our future work.
CA-SVM \cite{you2015svm} \cite{youdesign} used the divide-and-conquer approach for machine learning applications. 
The differences include: (1) CA-SVM is for classification while this work is for regression. 
(2) CA-SVM uses an iterative method, whereas we use a direct method. 
(3) The iterative method did not store the huge Kernel matrix, thus CA-SVM work has no consideration on the huge Kernel matrix. The scalability of iterative method is also limited.

%Lu et al 
Lu et al \cite{lu2013faster} designed a fast ridge regression algorithm, which reduced
the running time from $\Theta(n^2d)$ to $\Theta(log(n)nd)$. However, first,
this work made an unreasonable assumption: the number of features is much larger than
the number of samples ($d \gg n$). This is true for some datasets like the
webspam dataset \cite{regressiondata}, which has 350,000 training samples and each
sample has 16,609,143 features. However, on average only about 10 of these 16 million features
are nonzero. This can be processed efficiently by the sparse format like CSR (Compressed Sparse Row). Besides, this method only
works for the linear ridge regression algorithm rather than the kernel version,
which will lose non-linear space information.

%The most related work to this paper is Divide-and-Conquer Kernel Ridge Regression \cite{zhang2013divide},
%which is the serial version of DC-KRR. The basic idea is to divide the original data into $p$ parts,
%and then from each part construct a local smaller kernel matrix and finish the training process individually.
%DC-KRR is generally a fast serial algorithm, which reduces the memory requirement from
%$\Theta(n^2)$ to $\Theta(n^2/p^2)$. We finish a parallel implementation to justify it is also an efficient parallel approach.
%The idea of DC-KRR is to randomly and evenly divide the dataset into $p$ similar parts and get $p$ models. These $p$ models are {\bf similar} to each other because of the dividing algorithm. DC-KRR use an {\bf average} of these models, which has been shown better than the model generated by $1/p$ of the original dataset.
%The idea of BKRR2 or KKRR2 is to use a better partition algorithm to divide the datasets into $p$ different parts. There $p$ models are {\bf different} from each other because of the clustering approach. Either BKRR2 or KKRR2 are actually looking the {\bf best} model from these $p$ models.
%By doing so, BKRR2 achieves a much better accuracy than DC-KRR and a higher speed by using the same hardware. KKRR2 even achieves higher accuracy than the original method. Fig. \ref{fig:cakrr_dckrr_diff} summarizes the difference between BKRR2 and DC-KRR.

\section{Conclusion}
Due to a $\Theta(n^2)$ memory requirement and $\Theta(n^3)$ arithmetic operations,
KRR is prohibitive for large-scale machine learning datasets when $n$ is very large.
The weak scaling of KRR is problematic because the total memory required will grow like $\Theta(p)$ per processor,
and the total flops will grow like $\Theta(p^2)$ per processor.
The reason why BKRR2 is more scalable than DKRR is that it removes all the communication in the training part and reduces the computation cost from $\Theta(n^3/p)$ to $\Theta((n/p)^3)$.
The reason why BKRR2 is more accurate than DC-KRR is that it creates $p$ different models and selects the best one.
Compared to DKRR, BKRR2 improves the weak scaling efficiency from 0.32\% to 92\% and achieves 723$\times$ speedup for the same accuracy on the same 1536 processors.
Compared to DC-KRR, BKRR2 achieves much higher accuracy with faster speed. DC-KRR can never get the best accuracy achieved by BKRR2.
When we increase \# samples from 8k to 128k and \# processors from 96 to 1536, BKRR2 reduces the MSE from 93 to 14.7, which solves the poor accuracy weak-scaling problem of DC-KRR (MSE only decreases from 89 to 81).
KKRR2 achieves 591$\times$ speedup over DKRR for the same accuracy by using the same data and hardware. Based on a variety of datasets used in this paper, we observe that KKRR2 is the most accurate method. BKRR2 is the most practical algorithm that balances the speed and accuracy.
In conclusion, BKRR2 and KKRR2 are accurate, fast, and scalable methods.% The source code can be downloaded at \cite{cakrrcode}.

\section{Acknowledgement}
Yang You was supported by the U.S. DOE Office of Science, Office of Advanced Scientific Computing Research under Award Numbers DE-SC0008700 and AC02-05CH11231.
Cho-Jui Hsieh acknowledges the support of NSF via IIS-1719097
%Computing Research, Applied Mathematics program under Award Number DE-SC0010200
%We'd like to thank Prof. Inderjit Dhillon at UT Austin, Prof. Le Song at Georgia Tech, Prof. Martin Wainwright at UC Berkeley, and Dr. Yuchen Zhang at Stanford University for their discussions with us.

%YY and JD are supported by the U.S. Department of Energy Office of Science, Office of Advanced Scientific
%Computing Research, Applied Mathematics program under Award Number DE-SC0010200; by the
%U.S. Department of Energy Office of Science, Office of Advanced Scientific Computing Research
%under Award Numbers DE-SC0008700 and AC02-05CH11231; by DARPA Award Number HR0011-
%12-2-0016, Intel, Google, HP, Huawei, LGE, Nokia, NVIDIA, Oracle and S Samsung, Mathworks
%and Cray. CJH thanks the XSEDE and Nvidia support.

\bibliographystyle{ACM-Reference-Format}
\bibliography{sigproc}

\end{document}